\documentclass{PoS}
\newcommand{\sigmav}{$\langle\sigma v\rangle$ }

\def\lsim{\mathrel{\rlap{\lower4pt\hbox{\hskip1pt$\sim$}}
    \raise1pt\hbox{$<$}}}                
\def\gsim{\mathrel{\rlap{\lower4pt\hbox{\hskip1pt$\sim$}}
    \raise1pt\hbox{$>$}}}                

\title{AMS-02 electrons and positrons: astrophysical interpretation and Dark Matter constraints.}

\ShortTitle{AMS-02 leptonic data: PWN VS DM }

\author{\speaker{Di Mauro Mattia}
\\
        Universit\`a Degli Studi di Torino\\
        E-mail: \email{mattia.dimauro@to.infn.it}}

\author{Vittino Andrea\\
        Universit\`a Degli Studi di Torino\\
        E-mail: \email{vittino@to.infn.it}}

\abstract{We present here a quantitative analysis of the recent AMS-02 data with the purpose of investigating the interplay between astrophysical sources and Dark Matter in their interpretation. First, we show that AMS-02 leptonic measurements are in a remarkably good agreement with the hypothesis that all electrons and positrons are the outcome of primary or secondary astrophysical processes. Then, we add Dark Matter to the picture, in order to establish which are the informations on its annihilation cross section (or lifetime) that can be inferred by fitting AMS-02 data within a scenario in which Dark Matter and astrophysical sources jointly contribute to the different leptonic observables. In particular, by performing a Markov Chain Monte Carlo sampling of the parameters space of the theory, we attempt at characterizing the significance of a possible Dark Matter contribution to the observed data and we derive robust upper limits on the Dark Matter annihilation/decay rate.  }

\FullConference{The 34th International Cosmic Ray Conference,\\
		30 July- 6 August, 2015\\
		The Hague, The Netherlands}

\begin{document}

\section{Introduction}
The Alpha Magnetic Spectrometer (AMS-02) onboard the International Space Station is the state-of-the-art cosmic ray (CR) detector. After 30 months of data acquisition, the AMS-02 Collaboration has recently released the energy spectra in the 0.50$-$500 GeV range of the electron ($e^-$), positron ($e^+$) and inclusive ($e^+ + e^-$) fluxes and of the positron fraction ($e^+/(e^+ + e^-)$) \cite{2014PhRvL.113l1102A,2014PhRvL.113v1102A,2014PhRvL.11312}.

For energies larger than 30 GeV, the $e^+ + e^-$ energy spectrum can be perfectly reproduced by a power-law $dN/dE \propto E^{-\alpha}$ with $\alpha=−3.170 \pm 0.011$. In the same energy range, an analogous power-law behaviour, but with a slope of about $-2.8$,  characterises the positron spectrum. 
The difference in the slopes of the two spectra results in an increasing positron fraction for energies larger than $\approx 10$ GeV. This rise in the $e^+/(e^+ + e^-)$ was first indicated by HEAT \cite{1997ApJ...482L.191B}, then measured by PAMELA \cite{2009Natur.458..607A} and {\it Fermi}-LAT \cite{2012PhRvL.108a1103A} and finally confirmed with an unprecedented level of precision by AMS-02 \cite{2013PhRvL.110n1102A}.

The behaviour of the positron fraction in the high energy window is inconsistent with the traditional hypothesis that the positron flux is of purely secondary origin, {\it i.e.}, generated by spallation reactions of primary cosmic rays (mainly protons and helium nuclei) impinging on the nuclei that populate the interstellar medium (mostly hydrogen and helium).
In fact, the energy spectrum of secondary $e^+$ has a power-law slope of about $-3.5$ \cite{Delahaye:2008ua} which would give rise to a positron fraction that decreases with energy with a $-0.3$ slope, in strong tension with the observed slope which is about $+0.3$. 
This excess has been interpreted in terms of an extra-component of primary $e^+$ from Pulsar Wind Nebulae (PWNe) (e.g. \cite{Hooper:2008kg}), Supernova Remnants (SNRs) (e.g. \cite{2009PhRvL.103e1104B}) or as a possible hint of annihilating or decaying dark matter (DM) particles (e.g. \cite{Cirelli:2008pk}).

In Ref.~\cite{DiMauro:2014iia} we have built a model of the $e^{\pm}$ emission from PWNe and SNRs and we have illustrated how these contributions can be invoked, together with the secondary emission, in order to explain AMS-02 data in their whole energy range. In Ref.~\cite{DiMauro:2015jxa} we, on the other hand, focus on the high energy region of the spectra, namely $E\geq10$ GeV, with the purpose of inferring possible informations on the annihilation/decay rate of a particular class of DM candidates, that is Weakly Interactive Massive Particles (WIMPs). 


\section{Astrophysical interpretation of AMS-02 data}\label{sec:astro}
Cosmic electron and positron of astrophysical origin can be the outcome of a primary or secondary process. Primary electrons and positrons are emitted by sources such as PWNe and SNRs, while, as already pointed out, 
the secondary production refers to those particles that are the product of spallation reactions involving primary CRs and the nuclei of the interstellar medium.
In the next paragraphs we briefly summarize the basic features of these emission mechanisms in connection with the interpretation of AMS-02 data. 
We refer to \cite{DiMauro:2014iia,DiMauro:2015jxa} for the full details of the analysis.

\subsection{Primary emission from Pulsar Wind Nebulae}
A PWN is powered by a rapidly spinning neutron star which hosts, on its surface, a strong magnetic field that possesses an intensity in the range of $10^{7} - 10^{12}$ G. Because of the large intensity of this magnetic field, charged particles can be torn away from the surface of the star. These particles then multiply through electromagnetic cascade effects and end up in forming the {\it nebula}, that is a wind of particle and antiparticle pairs (e.g. \cite{Ruderman:1975ju}) that surrounds the pulsar.
Particles of the nebula are accelerated up to very high energies through shock mechanisms that originate from the interaction of the nebula with the slow ejecta of the SNR that is associated with the pulsar. 
It is typically assumed that this acceleration process lasts for around 50 kyr. After this period, the nebula is disrupted and the particles are injected in the interstellar medium in a burst-like event. 

We evaluate the PWN contribution to the $e^{\pm}$ flux by following the prescriptions outlined in Ref.~\cite{DiMauro:2014iia}. Basically, we consider all the sources listed in the ATNF catalogue \cite{Manchester:2004bp} and we model the injection spectrum of each one of them as a power-law with an exponential cut-off. Under this assumption, the $e^{\pm}$ flux associated to each PWN is completely described by 6 parameters: the spin down luminosity $\dot{E}$, the age $T$, the distance $d$, the spectral index $\gamma_{\rm{PWN}}$, the cutoff energy $E_c$ and the efficiency $\eta$ of the conversion of the total spin-down energy $W_0$ into $e^+e^-$ pairs (we address the reader to Ref.~\cite{DiMauro:2014iia,DiMauro:2015jxa} for a detailed description of these parameters). 
 
We assume the same cutoff energy $E_c=2$ TeV for every PWN, while the parameters $\dot{E}$, $d$ and $T$ which are specific to each source, are fixed to the values reported in the ATNF catalogue. Finally, $\eta$ and $\gamma_{\rm{PWN}}$, which are assumed to be common to all the PWNe, are free parameters of the model and will be determined by fitting AMS-02 data.

\subsection{Primary emission from Supernova Remnants}
SNRs are believied to be the most powerful accelerator of CRs in the Galaxy. These sources are capable of accelerating electrons up to very high energies through the interaction of these particles with the non-relativistic expanding shock-waves that are activated by the star explosion. As for the PWNe, the  spectrum that is expected to be injected in the interstellar medium by SNRs can be described in terms of a power-law with an exponential cut-off.
Informations about the $e^{-}$ flux originated by SNRs can be inferred by assuming the radio flux $B_r^\nu$ measured at a specific frequency $\nu$ in the shock region to be the synchrotron emission from the electrons accelerated by the SNR. As widely discussed in Refs.~\cite{2010AA...524A..51D,DiMauro:2014iia} this flux depends on the intensity of the magnetic field $B$ in the remnant, the spectral index $\gamma_{\rm{SNR}}$, the normalization $Q_0$ of the $e^-$ spectrum, the age $T$ of the source and its 
distance $d$ from the observer. 

We divide the whole population of  Galactic SNRs into two categories based on their distance: the {\it near} SNRs are the ones for which $d \leq 3$ kpc, while the {\it far} SNRs are characterised by $d>3$ kpc. We take the parameters of the {\it near} SNRs from the Green catalogue \cite{2009BASI...37...45G}
; we allow for a free normalization of the flux generated by the Vela SNR, which is the dominant $e^-$ contributor among local sources. Clearly, this normalization reflects changes in the magnetic field $B_{\rm{Vela}}$. 
On the other hand, the {\it far} component is treated as an average population of sources that follow the radial profile derived in Ref.~\cite{2004IAUS..218..105L} and that share common values for $Q_0$ and $\gamma$, which are free parameters that will be determined in our fit to AMS-02 data.

\subsection{Secondary production}
As discussed in Ref.~\cite{Delahaye:2008ua}, a crucial step in computing the secondary contribution to the $e^\pm$ fluxes, is the evaluation of the fluxes of the incoming primary CRs $\Phi_{\rm CR}$. 
We address this point by performing a fit to the AMS-02 measurements of the proton and Helium energy spectra \cite{2015PhRvL.114q1103A,ams02helium}. In particular, we consider a fitting function given by $\Phi(R<R_{\rm{break}})=A \beta^{P}R^{-P_1}$ and $\Phi(R\geq R_{\rm{break}})=A \beta^{P}R_{\rm{break}}^{-P_1+P_2}R^{-P_2}$, where $R=pc/Ze$ is the rigidity of the nucleus of charge number $Z$ and momentum $p$. The parameters that we obtain from the fit are: $A=26700 \pm  500$ m$^{-2}$s$^{-1}$sr$^{-1}$(GeV/n)$^{-1}$, $P=7.2 \pm 0.4$ and  $P_1=2.877 \pm 0.004$, $P_2= 2.748 \pm 0.013$ and $R_{\rm{break}} = 220\pm22$ GV for the proton flux, and  $A= 4110\pm 80 $ m$^{-2}$s$^{-1}$sr$^{-1}$(GeV/n)$^{-1}$, $P= 3.5\pm 0.7$ and $P_1= 2.793 \pm 0.004$, $P_2= 2.689 \pm 0.013$ and $R_{\rm{break}} = 187 \pm18$ GV for the helium flux (and for a Fisk solar modulation potential of $700 \pm 22$ MV).

\subsection{Fit results}\label{sec:astrofit}

\begin{figure*}[t]
	\centering
	\includegraphics[width=0.49\columnwidth]{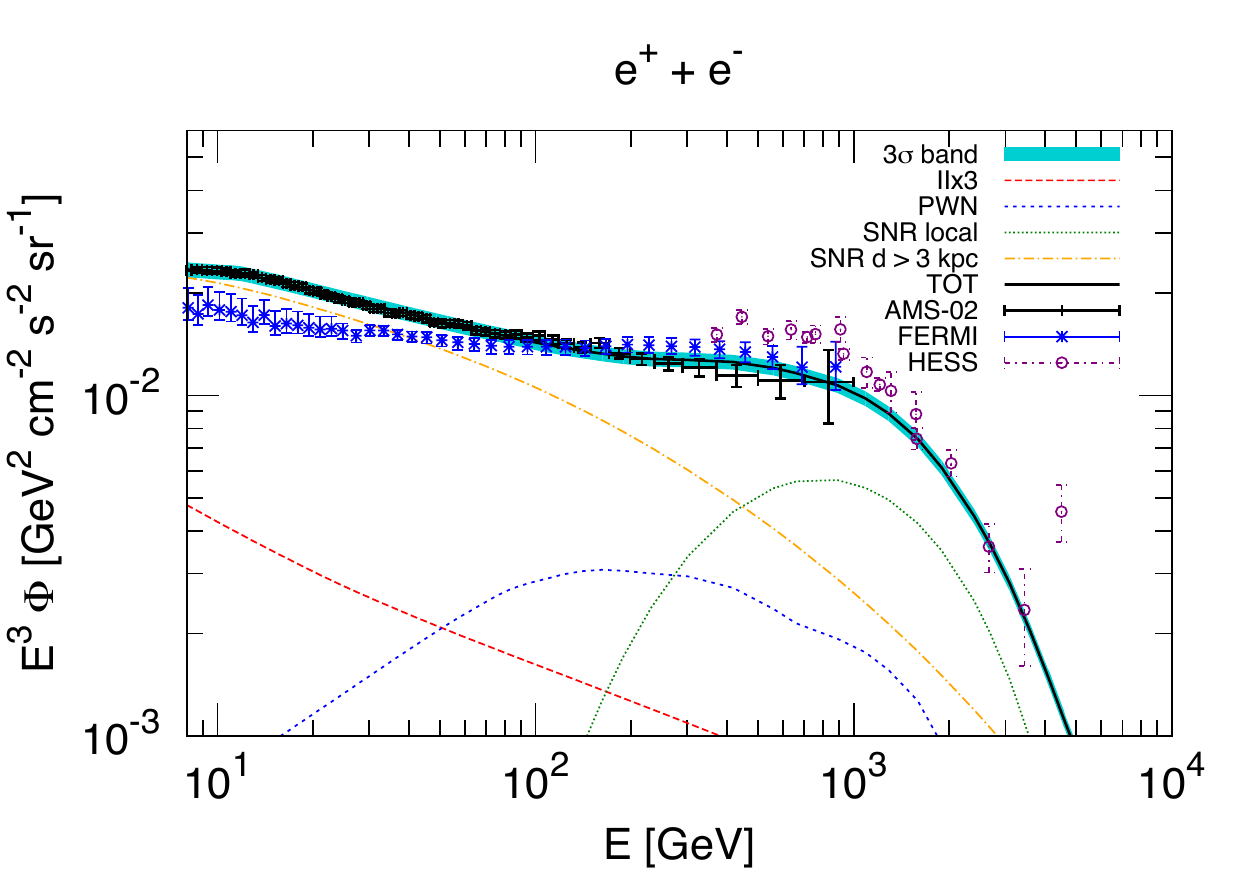}
	\includegraphics[width=0.49\columnwidth]{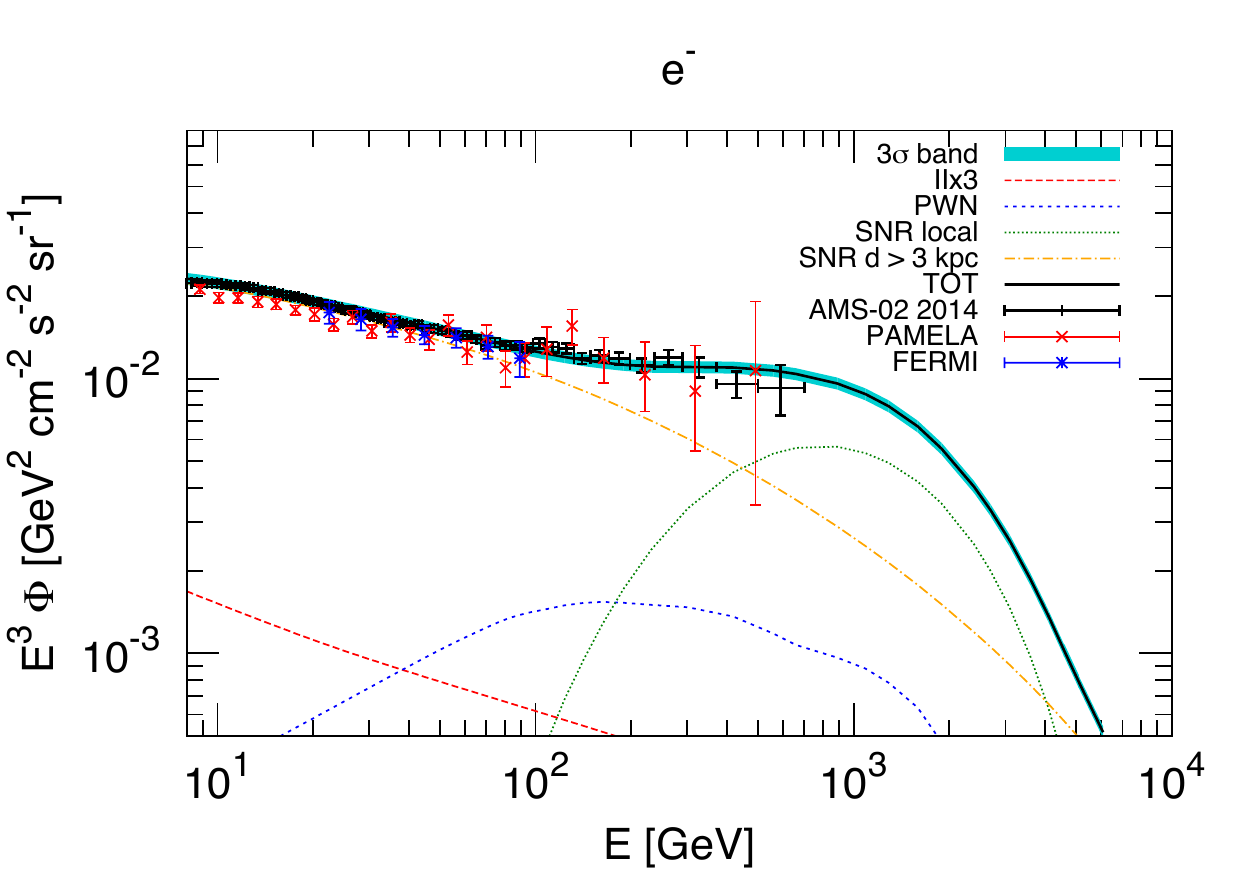}
	\includegraphics[width=0.49\columnwidth]{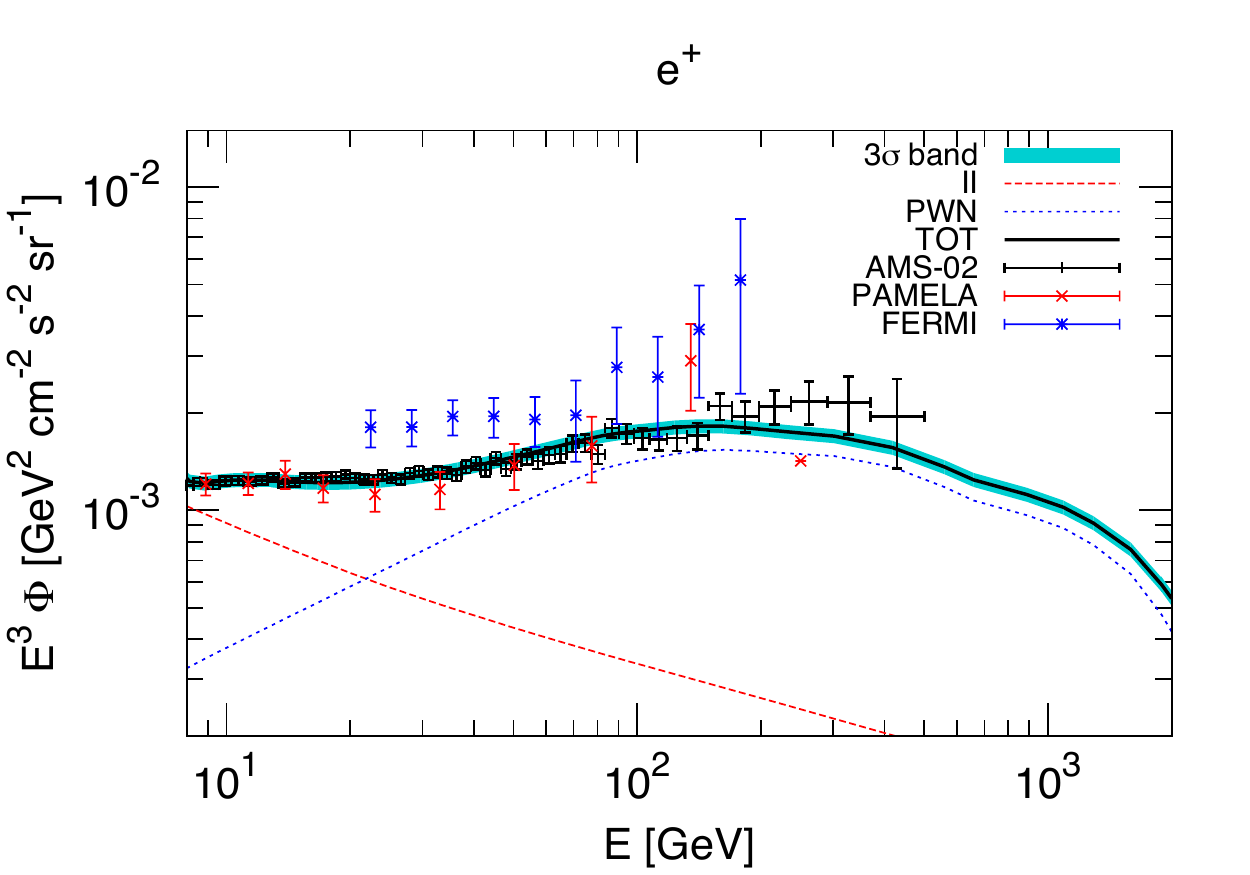}
	\includegraphics[width=0.49\columnwidth]{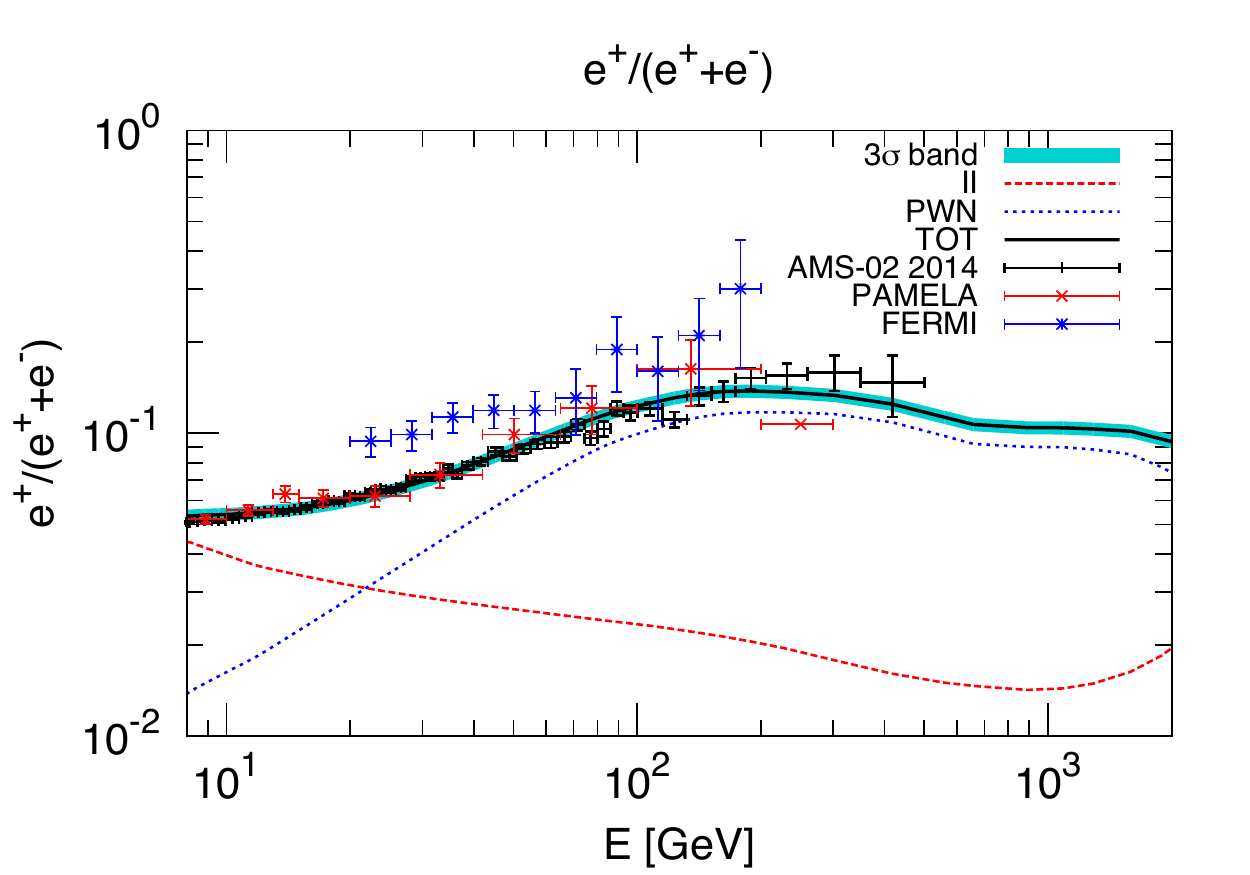}
\caption{The results of our fit to AMS-02 data of the $e^+ + e^-$ (top left), $e^-$ (top right),  $e^+$ (bottom left) and $e^+/(e^+ + e^-)$ (bottom right) energy spectra. In each panel, the solid black line represents the best fit results, and it is embedded in its 3$\sigma$ uncertainty band (cyan strip), while the different styles and colors of the other lines represent the various contributions from the component of the model: far SNRs (dot-dashed yellow), local SNRs (dotted green), PWNe (short dashed blue) and secondaries (long dashed red). Data from AMS-02 and HESS \cite{2008PhRvL.101z1104A} are also reported.}
\label{fig:astrofit} 
\end{figure*}

The electrons and positrons, that originate from the astrophysical sources discussed above, have to travel across the Galaxy and the heliosphere before they can reach the Earth. For the Galactic propagation we refer to the semi-analytical framework of the {\it two-zone diffusion model}, while to take into account the effects of solar modulation we adopt the {\it force field approximation}. The parameters related to the Galactic propagation are the ones of the MED propagation model \cite{2004PhRvD..69f3501D} while the Fisk potential $\phi$ of the solar modulation is a free parameter of the fit.

We perform a fit to the AMS-02 $e^-$, $e^- + e^+$, $e^+$ and $e^+/(e^-+e^+)$ energy spectra by means of a Markov Chain Monte-Carlo scan of the parameters space of our model performed with the CosmoMC package \cite{Lewis:2002ah}.
As discussed above, our model is characterised by 6 free parameters: $\phi$, $\gamma_{\rm{PWN}}$, $\eta$, $\gamma_{\rm{PWN}}$, $Q_0$ and $B_{\rm{Vela}}$.
The best-fit configuration that we find is associated to a reduced chi-squared $\chi^2/{\rm d.o.f.}=1.03$ with 191 data points and is characterised by the following values for the free parameters: $\phi = (0.14\pm0.07)$ GV, $\gamma_{\rm{PWN}}= 1.95^{+0.03}_{-0.02}$, $\eta= 3.68^{+0.0011}_{-0.0014}$, $\gamma_{\rm{SNR}}= 2.238^{+0.015}_{-0.013}$, $Q_0= (1.23^{+0.01}_{-0.03})\times 10^{50}$ GeV$^{-1}$ and $B_{\rm{Vela}}= (30\pm3)$ $\mu$G.
The best fit configuration and the $3\sigma$ uncertainty band are displayed in Fig.~\ref{fig:astrofit} for $e^-$, $e^- + e^+$, $e^+$ and $e^+/(e^-+e^+)$ AMS-02 data. The situation illustrated in the Figure and the low chi-squared that we find are a strong indication that our astrophysical model provides a remarkably good explanation of AMS-02 data.

\section{Dark matter constraints}\label{sec:dmul}
\begin{figure*}[t]
	\centering
	\includegraphics[width=0.49\columnwidth]{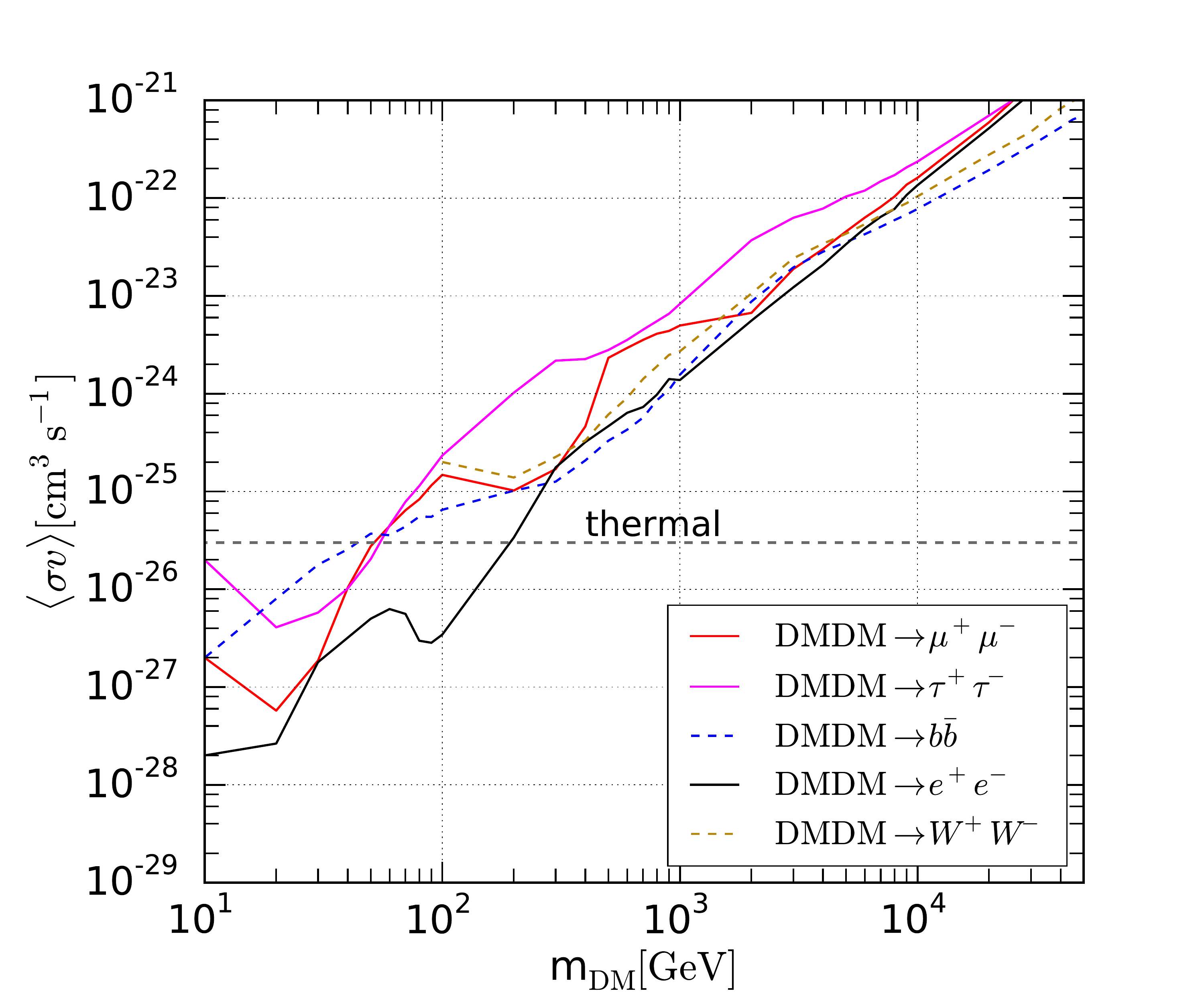}
	\includegraphics[width=0.49\columnwidth]{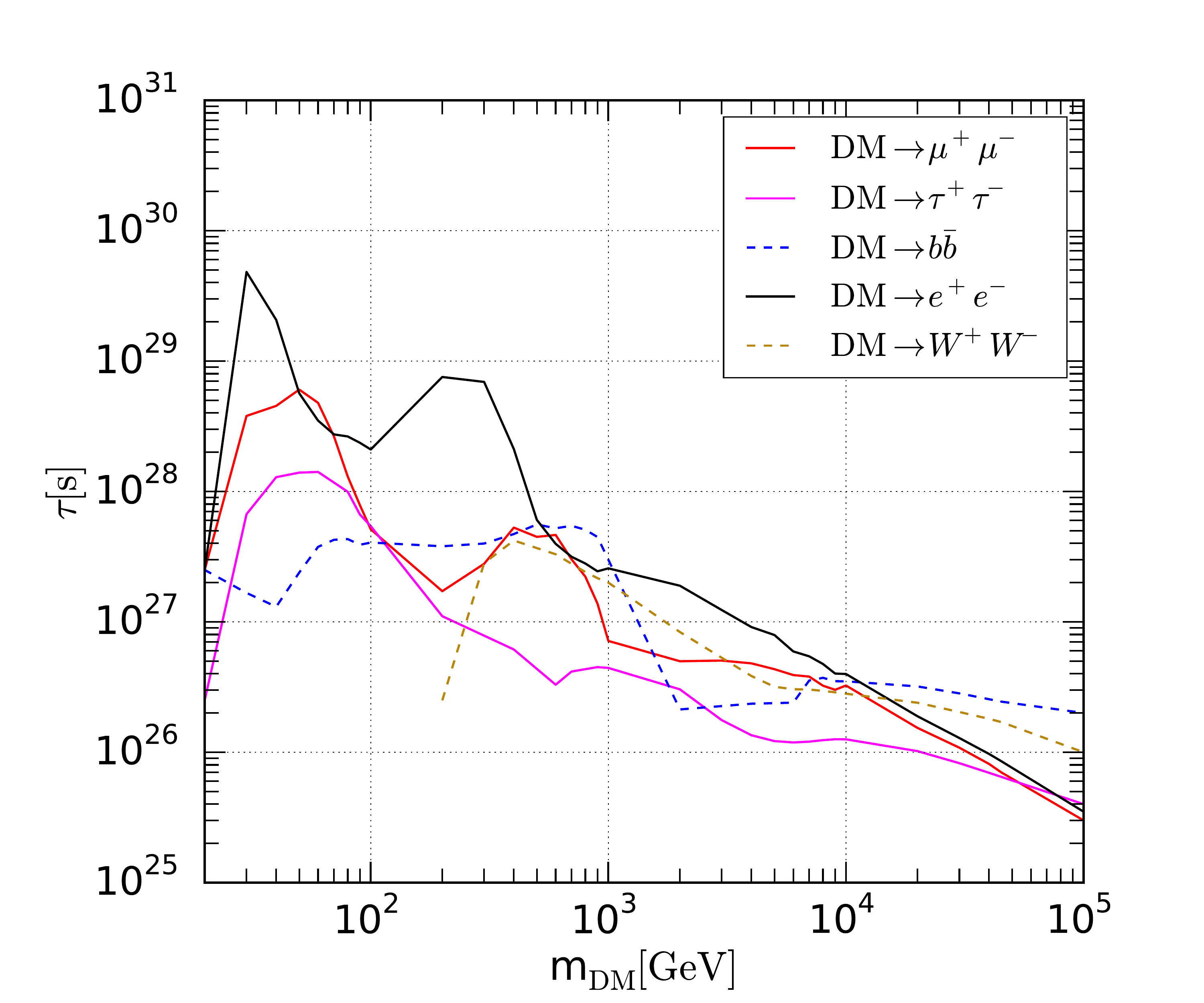}
\caption{The bounds on \sigmav (left panel) and $\tau$ (right panel) are reported for several DM annihilation/decay channels.}
\label{fig:bounds} 
\end{figure*}

In this Section, we expand the analysis described in Section~\ref{sec:astrofit} with the purpose of deriving robust constraints on the DM annihilation cross section \sigmav or lifetime $\tau$. We consider the 6 annihilation/decay channels: $DMDM\rightarrow$ ($DM \rightarrow$) $e^+e^-$, $\mu^+\mu^-$, $\tau^+\tau^-$, $b\bar{b}$, $t\bar{t}$ or $W^+W^-$. For each one of these channels, we determine the $e^{\pm}$ flux by following the approach described in Ref.~\cite{Delahaye:2007fr}: for the injected energy spectra $dN/dE$ we refer to Ref.~\cite{Cirelli:2010xx}, while we asssume for the DM distribution in the Galaxy a NFW profile and a local density of $0.4$ GeV/cm$^3$ (we have checked that our results and conclusions do not change significantly if other density profiles are assumed). 

The model that we use to fit AMS-02 data consists of the astrophysical contributions described in Sec.~\ref{sec:astrofit} with the addition of a DM component that is computed for fixed values of the DM mass. Therefore, the fit depends on 7 free parameters: the 6 aformentioned degrees of freedom that characterize the astrophysical sources plus the parameter quantifying the DM emission, that is \sigmav for the annihilating case or $\tau$ if we are considering decaying DM.  Within a purely {\it bayesian} approach, the upper limits on \sigmav and $\tau$ are identified as the values for which the area underlying the posterior distributions of these parameters correspond to the desired confidence level (in our case, about 0.955, that corresponds to $2\sigma$ bounds). 

The bounds that we obtain are displayed in Fig.~\ref{fig:bounds}; as it can be seen, they are particularly strong for the $e^+e^-$ and $\mu^+ \mu^-$ channels, for which, in the annihilating case, we are able to probe the thermal cross section for DM masses up to 200 and 70 GeV, respectively.

\section{Dark matter interpretation}\label{sec:dmbf}
\begin{figure*}[t]
	\centering
	\includegraphics[width=0.49\columnwidth]{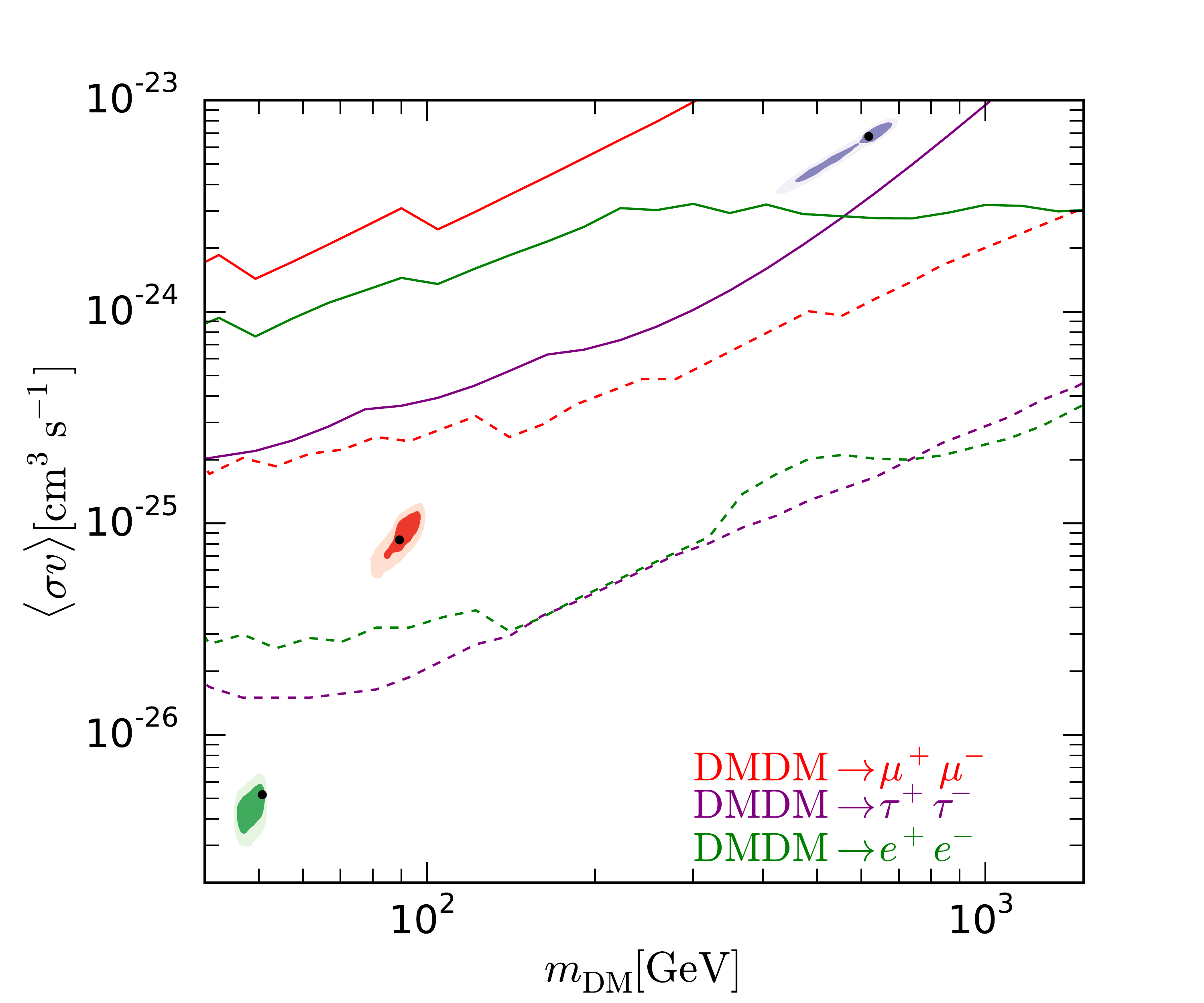}
	\includegraphics[width=0.49\columnwidth]{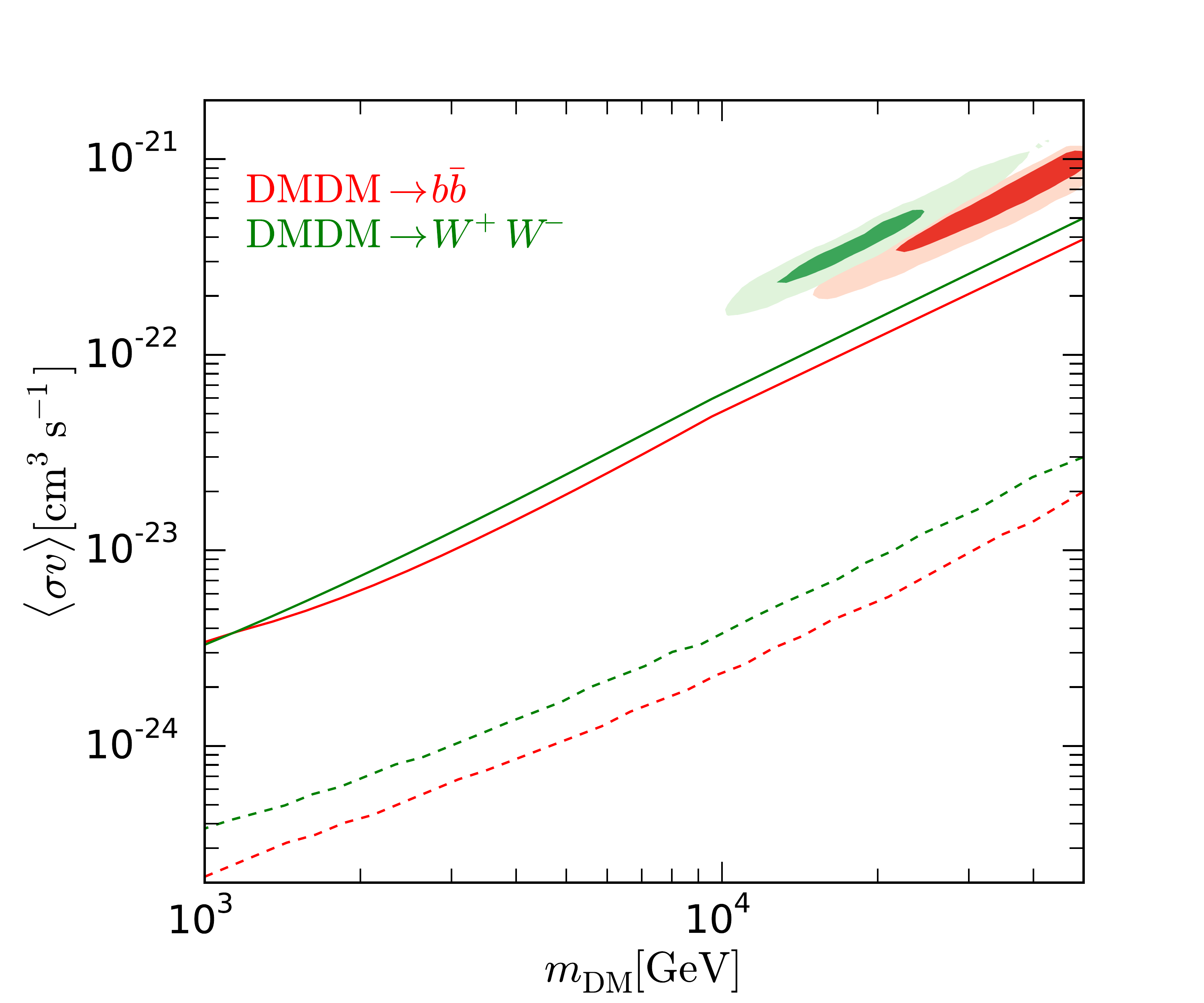}
\caption{Darker (lighter) shadings represent the $1\sigma$ and $2\sigma$ contour regions that are associated to the best fits to AMS-02 data for a DM particle annihilating into leptonic ({\it left panel}) and hadronic ({\it right panel}) channels. In the two panels, upper limits derived from $\gamma$-rays are shown, as discussed in the text. }
\label{fig:cpdm} 
\end{figure*}

The purpose of this Section is to investigate the possible presence of a DM signal hidden in the data. To accomplish this task, we once again perform a fit to AMS-02 data within a model that consists of the joint contribution from both primary astrophysical sources, secondary production and from DM annihilation. As already widely discussed, the astrophysical emission is determined by 6 free parameters, while two parameters, namely \sigmav (or $\tau$) and $m_{\rm{DM}}$,  regulate the entity of DM contribution (we remark that, following a different approach with respect to the one outlined in the previous Section, we treat $m_{\rm DM}$ as a free parameter of the fit). 

Fig.~\ref{fig:cpdm} shows the $1\sigma$ and $2\sigma$ contour regions in the $(m_{\rm DM},\langle\sigma v\rangle)$ plane for the annihilation channels under scrutiny (the contour for annihilation into $t\bar{t}$ is basically coincident with the one of the $b\bar{b}$ channel). Together with these best-fit configurations, the figure also reports for each annihilation channel the upper limit on \sigmav that can be derived from the study of the DM contribution to the {\it Isotropic Gamma Ray Background} (IGRB). In particular, solid lines represent {\it conservative} upper limits derived in Ref.~\cite{Calore:2013yia} by setting the $\gamma$-ray emission from astrophysical sources ({\it e.g.}  Active Galactic nuclei, pulsars and Star Forming Galaxies) to the expected minimum, while dashed lines denote {\it optimistic} constraints derived in Ref.~\cite{DiMauro:2015tfa} by means of a more refined multi-component fit.   

An important remark that has to be made is that {\it all} the annihilation channels are able to provide an agreement with AMS-02 data that is better than the one that characterize the astrophysical interpretation presented in Sec.~\ref{sec:astrofit}. At the same time, it is clear that only the $e^+e^-$ and $\mu^+ \mu^-$ channels are able {\it at the same time} to satisfy the $\gamma$-ray bounds. In particular, the $\mu^+ \mu^-$ is the most promising one, being characterised by a $\Delta \chi^2 \sim 49$ with respect to the astrophysical fit and meaning that our model prefers with a significance of $6.9 \sigma$ the presence of DM. The best-fit parameters for this channel are $m_{\rm DM}=89^{+22}_{-10}$ GeV and \sigmav$= 8^{+{8}}_{-3} \times 10^{-26}$ cm$^3$s$^{-1}$. 

\begin{figure*}[t]
	\centering
	\includegraphics[width=0.54\columnwidth]{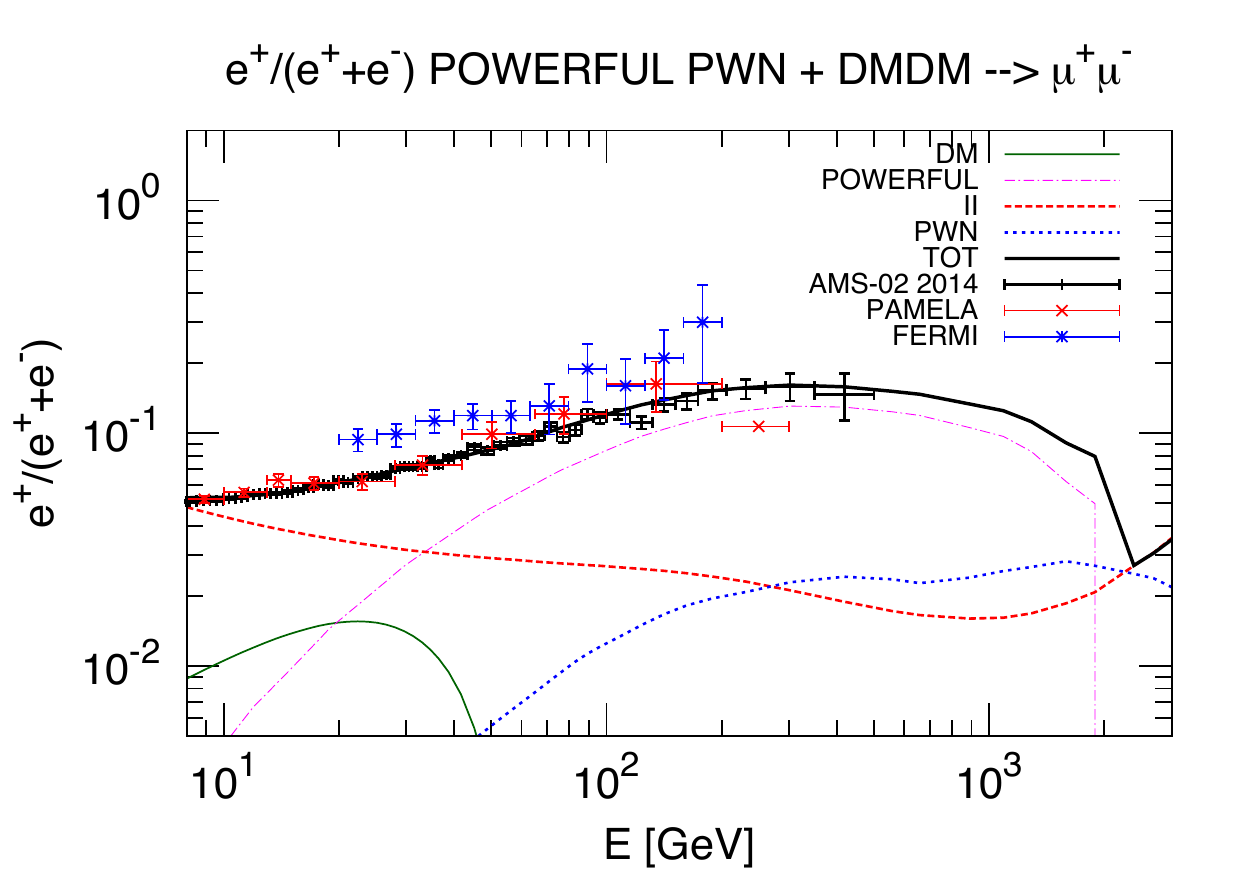}
	\includegraphics[width=0.45\columnwidth]{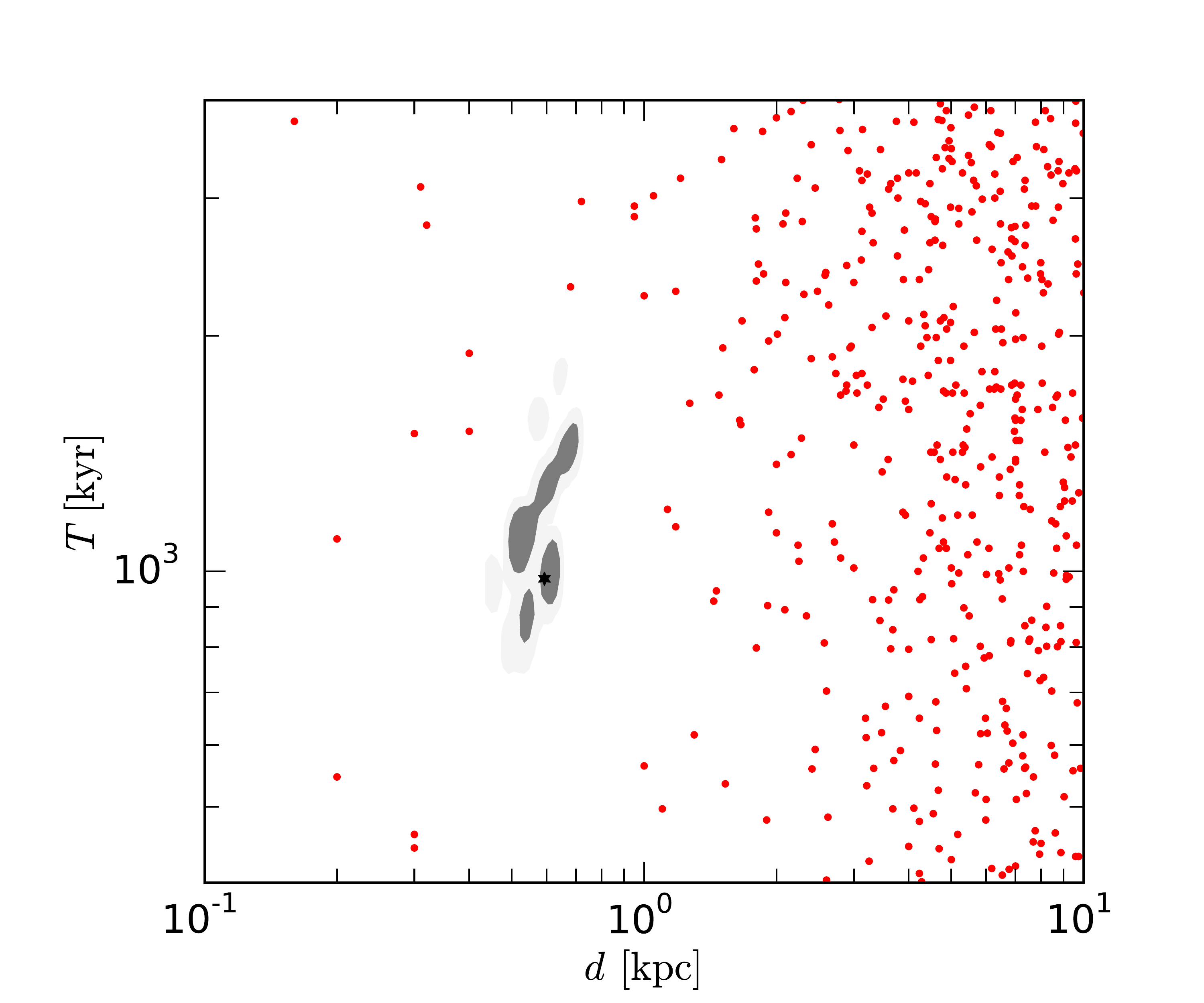}
\caption{In the {\it left panel}, the solid line reports the positron fraction as given by the jointly contribution of secondaries, PWNe and DM annihilating into the $\mu^+ \mu^-$ channel. The {\it right panel} shows the $1\sigma$ (darker shading) and $2\sigma$ (lighter shading) contour regions in the $(distance,age)$ plane associated to the additional PWN which, together with the sources of the ATNF catalog, can provide the best-fit to data.}
\label{fig:additional} 
\end{figure*}

To test this possible DM hint, we consider a refined version of the astrophysical model described in Sec.~\ref{sec:astrofit} in which the efficiencies of the five most powerful PWNe of the ATNF catalogue are left free to vary in the fitting procedure. These five pulsars which can be found by means of the ranking algorithm described in Ref.~\cite{DiMauro:2014iia}, are Geminga, Monogem, J2043+2740, J0538+2817 and B1742-30. If we fit the data without DM, the chi squared is 146, while the best-fit with the addition of DM is 122 in the case of $\mu^{+}\mu^{-}$ and 133 for $\tau^{+}\tau^{-}$ annihilation channel. It is worth noticing that, even within this model that allows more freedom to the astrophysical component, a DM annihilating into the $\mu^+ \mu^-$ or $\tau^{+}\tau^{-}$ channel is able to provide a better fit to AMS-02 data. 
In fact, we obtain $\Delta \chi^2 \sim 24$ for $\mu^{+}\mu^{-}$ and $\sim 13$ for $\tau^{+}\tau^{-}$ annihilation channel associated to a significance of DM, with respect to our astrophysical model, of 4.5 and 3.2$\sigma$ respectively.
The DM candidates has for $\mu^{+}\mu^{-}$ channel $m_{\rm DM}=51^{+10}_{-3}$ GeV and \sigmav$= 2.8^{+1.1}_{-0.5} \times 10^{-26}$ cm$^3$s$^{-1}$ while for $\tau^{+}\tau^{-}$ channel $m_{\rm DM}=140^{+7}_{-7}$ GeV and \sigmav$= 4.3^{+16.0}_{-4.2} \times 10^{-26}$ cm$^3$s$^{-1}$.
Therefore, the $\mu^{+}\mu^{-}$ DM candidate is fully consistent with $\gamma$-ray upper limits shown in Fig.~\ref{fig:cpdm} while the $\tau^{+}\tau^{-}$ channel candidate is in a certain tension with the {\it optimistic} $\gamma$-ray  bounds.
 \\
The contribution of the $\mu^{+}\mu^{-}$ DM candidate to the positron fraction measured by AMS-02 is shown in the left panel of Fig.~\ref{fig:additional}. It clearly stands out that the reason why the addition of a relatively light DM annihilating into leptons helps in improving the fit is that the positron flux that stems from this source provides a sizeable contribution at those intermediate energies at which the contribution from the PWNe of the ATNF catalog is relatively weak. In other words, DM helps in ``filling the gap" where the secondary flux connects to the contribution from pulsars. 
\\
With this being said, one can easily conclude that what has been presented until now as a possible hint of DM can also be seen as a mere consequence of the fact that the sources listed in the ATNF catalog constitute only a limited sample of the total number of PWNe in the sky. We illustrate this point by fitting AMS-02 data within the astrophysical model described in Sec.~\ref{sec:astrofit} but with an {\it additional PWN}, whose distance, age and total emitted energy in $e^{\pm}$ pairs being free parameters of the fit. The best-fit configuration is given by a chi-square value of 146 ($\chi^2/{\rm d.o.f.}=0.70$) a source with a distance $0.59^{+0.11}_{-0.15}$ kpc and age $980^{+820}_{-210}$ kyr and is characterized by a $\Delta \chi^2$ which is comparable to the one given by DM.

\section{Conclusions}
In this work we have provided a quantitative characterization of the electron and positron data measured by the AMS-02 experiment. In particular, confirming the results already reported in Ref.~\cite{DiMauro:2014iia}, we have shown that the emission from primary and secondary astrophysical sources provide a perfectly viable interpretation of the experimental measurements. As a consequence, these leptonic observables can be used to impose strong constraints to the emission from DM annihilation or decay: in particular, for the leptonic channels, the bounds that we have obtained here, by means of a purely bayesian approach, are about as strong as the constraints that can be derived from $\gamma$-rays investigations. Finally, we have illustrated how a contribution from DM annihilating into $\mu^{+}\mu^{-}$ with a mass $m_{\rm DM}=50$ GeV and an annihilation cross section \sigmav$= 2.8 \times 10^{-26}$ cm$^3$s$^{-1}$, which is thus fully compatible with $\gamma$-ray upper limits, can improve the fit  up to a 4.5$\sigma$ C.L. with respect to the purely astrophysical model. 
However, we have also illustrated how the addition of a pulsar with an age of about 1000 kyr and located at a distance of 700 pc from Earth could provide a comparably good fit of the experimental data even if a pulsar with such characteristics is not present in the ATNF catalog. Therefore, we conclude by saying that both DM and astrophysical sources are perfectly viable interpretations of AMS-02 measurements: in order to discriminate between these two alternatives, it will be crucial to dedicate future efforts on dedicated studies of the $e^{\pm}$ emission from PWNe.

\bibliography{skeleton_revised}

\end{document}